\documentclass[conference]{IEEEtran}
\IEEEoverridecommandlockouts
\usepackage{cite}
\usepackage{amsmath,amssymb,amsfonts}
\usepackage{graphicx,xcolor,lipsum,multicol}
\usepackage{textcomp}
\usepackage{changes}
\usepackage{comment}
\usepackage{algorithm}
\usepackage{algpseudocode}
\usepackage{xcolor} 
\usepackage{balance} 

\def\BibTeX{{\rm B\kern-.05em{\sc i\kern-.025em b}\kern-.08em
    T\kern-.1667em\lower.7ex\hbox{E }\kern-.125emX}}

\usepackage{geometry}
\newtheorem{proposition}{Proposition}
\newtheorem{corollary}{Corollary}
\newtheorem{remark}{Remark}

\begin{document}

\algnewcommand{\Initialize}{\Statex \textbf{Initialization:}}
\algnewcommand{\Iteration}{\Statex \textbf{Iteration:}}
\algnewcommand{\Output}{\Statex \textbf{Output:}}

\title{Joint Transmit Power and Reflection Coefficient Optimization for Underlay Cognitive Radio Ambient Backscatter Communication}

\author{\IEEEauthorblockN{Chabalala S. Chabalala\IEEEauthorrefmark{1},~\IEEEmembership{Member,~IEEE}, Peter Odero Akuon\IEEEauthorrefmark{1}, and
Zhen Gao\IEEEauthorrefmark{3}}
\IEEEauthorblockA{\IEEEauthorrefmark{1}School of Electrical \& Information Engineering,
University of the Witwatersrand, Johannesburg, South Africa}
\IEEEauthorblockA{\IEEEauthorrefmark{2}Department of Electrical \& Information Engineering, University of Nairobi, Nairobi, Kenya}
\IEEEauthorblockA{\IEEEauthorrefmark{3}Advanced Research Institute of Multidisciplinary Science, Beijing Institute of Technology, Beijing, China}
Email: \{chabalala.chabalala@wits.ac.za\IEEEauthorrefmark{1}, akuon@uonbi.ac.ke\IEEEauthorrefmark{2},
gaozhen16@bit.edu.cn\IEEEauthorrefmark{3}\}
}%

\maketitle

\begin{abstract}

The integration of cognitive radio (CR) technology with backscatter communication offers a promising solution for developing spectrum-efficient and low-power wireless networks. In this paper, we investigate the performance of an underlay cognitive radio ambient backscatter communication (CR-AmBC) system operating under primary network (PN) interference constraints. The objective is to maximize the performance of the CR-AmBC system while protecting the PN from interference. Differing from prior work, our formulation jointly accounts for: (i) interference from the PN on the CR receivers, (ii) interference from the backscatter device (BD) on the PN receiver, and (iii) imperfect successive interference cancellation (i-SIC) at the secondary destination. We propose a joint transmit power and reflection coefficient optimization framework that incorporates the minimum data rate requirements, the PN interference constraint, and i-SIC for decoding the BD signal. The nonconvex problem is decomposed into two subproblems solved via alternating optimization: (a) a closed-form expression for the feasible transmit power given a fixed reflection coefficient, and (b) an efficient one-dimensional search for the reflection coefficient. We prove that the power update is optimal, reducing the joint problem to a one-dimensional search, with monotone convergence and $\mathcal{O}(KN_\beta)$ complexity for $K$ iterations and $N_\beta$ search points. We present a comprehensive simulation-based performance evaluation and the results demonstrate that the proposed joint transmission power and reflection coefficient optimization provides an effective strategy to maximize the performance of the CR-AmBC system without violating the PN protection constraints.

\end{abstract}

\begin{IEEEkeywords}
Ambient backscatter communication (AmBC), cognitive radio networks (CRNs), interference management, outage performance, power reflection coefficient.
\end{IEEEkeywords}


\section{Introduction} \label{introduction}
The sixth generation (6G) technology is expected to support a wide range of emerging network applications, including massive machine-type communication (mMTC) services \cite{Mahmood2020}. In mMTC services, a large number of low-cost devices are envisioned for various Internet-of-Things (IoT) applications \cite{Nguyen2022}, \cite{Li2022}. Energy efficiency (EE) is a major requirement in fifth-generation (5G) networks and remains a primary requirement in the expected 6G wireless systems \cite{Hossfeld2024}. For example, the 3rd Generation Partnership Project (3GPP) new radio (NR) deployments are estimated to be more than four times as energy-efficient \cite{Alnahari2025}, \cite{Azzino2024}. In recent years, advances in electronics and wireless technologies have driven an increase in network applications and services, as well as the number of connected devices. Despite the growing demand for new services and applications, the available radio spectrum remains a finite resource \cite{Trichias2024}. Spectrum sharing is therefore a crucial enabling technique to address spectrum scarcity, facilitated by cognitive radio (CR) technology, which allows the coexistence of primary networks (PNs) and cognitive radio networks (CRNs) through dynamic spectrum access \cite{Eltanab2017}, \cite{Ahmad2020}. In underlay transmissions, primary users (PUs) and secondary users (SUs) perform concurrent transmissions using the same spectrum resources \cite{Cai2020}. In this context, interference management for SU transmissions are essential to protect PU transmissions.

Ambient backscatter communication (AmBC) is one of the promising techniques to address the challenge of ultra-low-power and self-sustainable IoT networks. In AmBC, devices do not generate radio frequency (RF) signals but instead utilize existing RF signals from the surrounding environment \cite{Huynh2023}, \cite{Zhang2019}, \cite{Van2018}, \cite{Raghavendra2026survey}. As a result, AmBC devices are low-cost, low-complexity, and require only a small amount of power, which can be harvested from ubiquitous RF signals in the environment \cite{Zhang2019}, \cite{Alnahari2025}, \cite{Liang2020}. Its practical relevance is underlined by the 3GPP Ambient IoT studies \cite{3GPP38769}, \cite{Kim2026ambient}, \cite{Famaey2026survey}, in which battery-less devices reuse the signals of existing networks, making the interference exchanged with the incumbent network a deployment issue. The authors in \cite{Badarneh2022} investigated the performance of CRN with backscatter under an adaptive transmit power strategy to protect the PN. However, their analysis did not take into account the interference caused by the PN. In \cite{Zhang2023}, the authors studied the outage performance of an underlay cognitive radio non-orthogonal multiple access (CR-NOMA) backscatter system, where the CRN transmitter adaptively adjusts its transmit power to keep the interference at the PN receiver below the interference threshold, but they did not consider interference from the PN transmitter at the CRN receivers. In \cite{Thenua2024}, \cite{Zhuang2020}, the authors investigated sum-rate maximization of an underlay uplink CR-NOMA system with multiple backscatter devices (BDs), optimizing the power reflection coefficients (PRCs) of multiple BDs while meeting the interference constraints at the PN receiver. In \cite{Jee2023}, the authors derived closed-form expressions for non-outage probabilities of the PN, CRN receiver, and BD transmissions. From outage expressions, the optimal transmit power and reflection coefficients were obtained. More recent designs for cognitive and symbiotic backscatter networks \cite{Xu2024ris}, \cite{Rezaei2024noma} rely on iterative solvers such as semidefinite relaxation (SDR) and successive convex approximation (SCA).

However, due to the complexity of the outage expressions, the optimal parameters were derived based on simplified approximations. Notwithstanding, most of the existing works in the literature have not investigated the impact of PU transmissions on CRN and BD data, and the effect of BD on the PN receiver has also been largely overlooked \cite{Zhang2019}, \cite{Badarneh2022}, \cite{Zhang2023}, \cite{Thenua2024}, \cite{Zhuang2020}, \cite{Jee2023}. The main distinction of the presented work is that we simultaneously model interference from the PN onto the CR receivers and the reflected interference from the BD onto the PN receiver, include imperfect successive interference cancellation (i-SIC) at the secondary destination, and derive explicit feasibility conditions for the joint optimization of \( (P_s,\beta)\). Furthermore, we show that this structure admits an optimal closed-form power update, so that the joint problem is solved by a one-dimensional search with provable convergence and low complexity. The main contributions are summarized as follows:
 
\begin{itemize}
    \item We formulate a joint transmit power and power-reflection coefficient optimization for an underlay cognitive radio ambient backscatter communication (CR-AmBC) system that explicitly models (i) interference from the PU transmitter on both the secondary receiver and BD receiver, and (ii) reflected interference from the BD onto the PN receiver. The model also incorporates i-SIC for BD decoding.
    
    \item We propose a joint transmit power and reflection coefficient optimization framework that decomposes the problem into two tractable subproblems: a closed-form solution or feasibility evaluation for the transmit power \(P_s\) given \(\beta\), and a single-variable search for \(\beta\). For the power subproblem we derive explicit feasibility conditions and closed-form expressions for \(P_{\min}(\beta)\) and \(P_{\max}(\beta)\). We prove that this power update is optimal, so the joint problem reduces to a one-dimensional search, and establish the convergence and \(\mathcal{O}(KN_\beta)\) complexity of the algorithm.

    \item We present a comprehensive performance evaluation using simulation experiments, which includes comparisons of the proposed joint-power and reflection coefficient technique with baselines for: (a) fixed-power and fixed-reflection coefficient, (b) fixed-power and optimal-reflection coefficient, (c) optimal-power and fixed-reflection coefficient of an underlay CR-AmBC system under PN interference constraints. Baselines (b) and (c) emulate the single-variable designs of \cite{Badarneh2022}, \cite{Zhang2023}, \cite{Thenua2024}, and practical design guidelines are derived.
\end{itemize}

The remainder of this paper is organized as follows: Section \ref{system} presents the system model. Section \ref{problem} presents the problem formulation, the proposed solution framework, and its convergence and complexity analysis; followed by Section \ref{results}, which presents the simulation results and discussion. Section \ref{conclutions} provides the main conclusions derived from the presented work.


\section{System Model} \label{system}

\begin{figure}[tbp!]
\centerline{\includegraphics[width=0.46\textwidth]{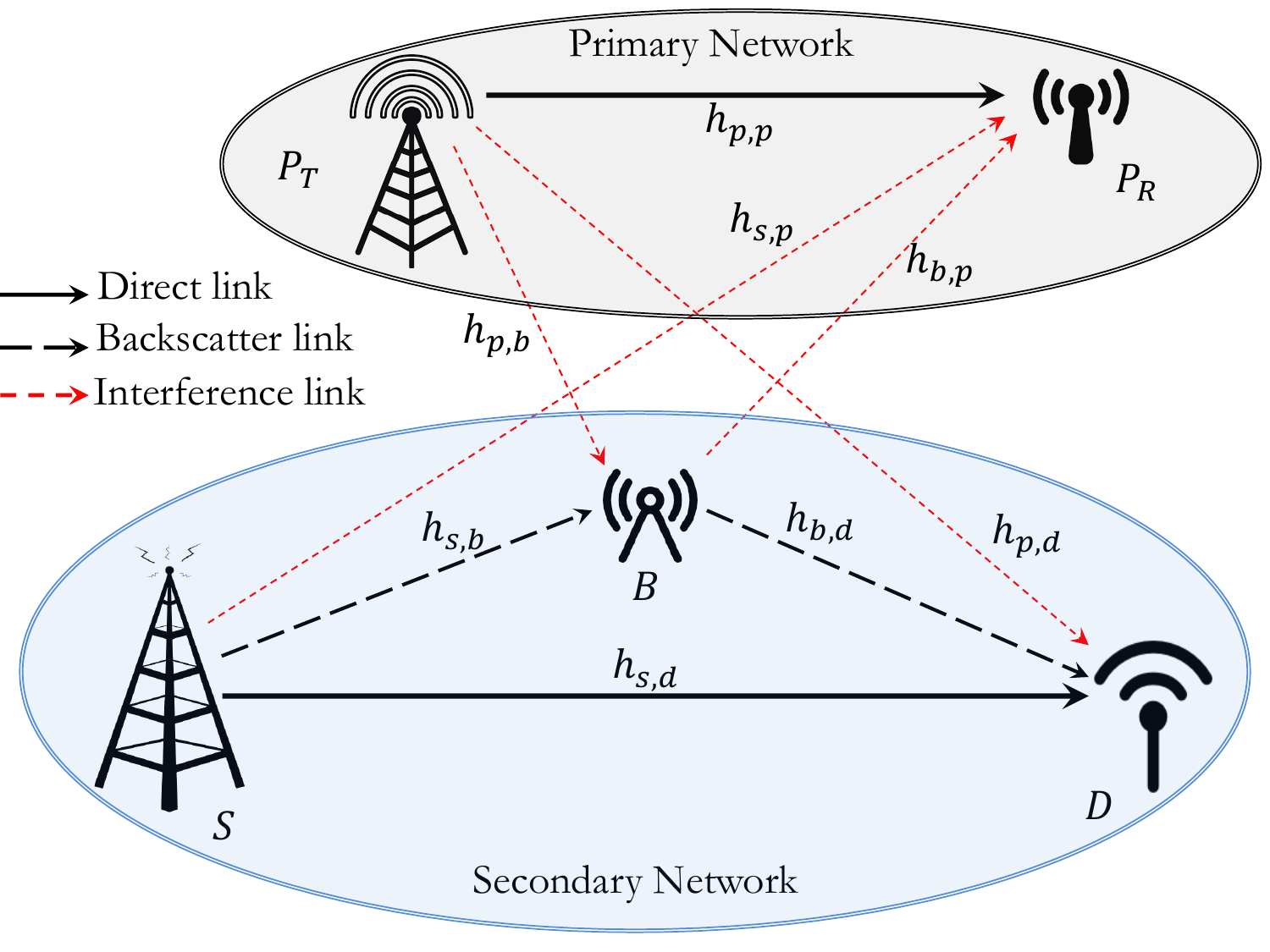}}
\caption{System model for underlay CRN with BD under PN interference, where the PN is subject to interference from the reflected signals.}
\label{fig:fig001}
\end{figure}


\subsection{Basic Assumptions} 

In this paper, we consider the system model presented in Fig. \ref{fig:fig001}, in which the PN comprises a transmitter $P_T$ and a receiver $P_R$, with a direct link through the channel $h_{p,p}$, as well as a CRN comprising a secondary transmitter (ST) node $S$, a backscatter device (BD) $B$, and a secondary destination (SD) node $D$. All channel gains are assumed to follow quasi-static Rayleigh fading and distance-dependent path loss, characterized by $h_{i,j} = g_{i,j} \sqrt{d_{i,j}^{-\alpha}}$ for all $i,j \in \{ P_T, P_R, S, D, B \}$, where $g_{i,j} \sim \mathcal{CN}(0,1)$ denotes the fading coefficient between nodes $i$ and $j$, $d_{i,j}$ is the distance, and $\alpha$ is the path loss exponent. The channel gains $|h_{i,j}|^2$ therefore follow an exponential distribution, i.e., $|h_{i,j}|^2 \sim \text{Exp}(\lambda_{i,j})$, where $\lambda_{i,j} \triangleq d_{i,j}^{-\alpha}$. All nodes have a single antenna, reflecting low-cost Ambient IoT devices \cite{Kim2026ambient}. We assume that the instantaneous channel state information (CSI) of each quasi-static block is available at nodes $S$ and $B$, which in practice, can be obtained by employing channel estimation techniques \cite{Thenua2024}, \cite{Jee2023}, and the optimization is performed once per block; imperfect or statistical CSI \cite{Xu2024ris} is left for future work.


\subsection{Signal Model} \label{signal}
Suppose $S$ transmits the signal $x_s$ with $\mathbb{E}{|x_s|^2} = 1$ to $D$ through $h_{s,d}$ ($S \rightarrow D$). The backscatter node re-modulates and reflects the received signals. The signal received at $B$ has two components from $S$ and $P_T$: $y_B = \sqrt{P_s} h_{s,b} x_s + \sqrt{P_p} h_{p,b} x_p$. There are four components of the received signal at $P_R$. The received signal at $P_R$ is then given by:
\begin{equation} 
    \begin{aligned}
    {y_{P}} &= \underbrace{\sqrt {P_p} {h_{p,p}}{x_{p}}}_{\text{$P_T$ direct}} + \underbrace{\sqrt {P_s} {h_{s,p}}{x_{s}}}_{\text{$S$ direct}} 
    + \underbrace{\sqrt {{\beta} P_p} {h_{p,b}}{h_{b,p}}{x_{p}}{x_{b}}}_{\text{$P_T$ backscatter}} \\
    &+ \underbrace{\sqrt {{\beta} P_s} {h_{s,b}}{h_{b,p}}{x_{s}}{x_{b}}}_{\text{$S$ backscatter}} + \underbrace{{n_{p}}}_{\text{noise}},
    \end{aligned}
\end{equation} 
where $n_p \sim \mathcal{CN}(0,\sigma_p^2)$ is the additive white Gaussian noise (AWGN) with zero mean and variance $\sigma_p^2$ at $P_R$, $P_t$ is the transmit power of source $\forall t \in \{s, p \}$, $\beta \in [0,1]$ is the power reflection coefficient and efficiency at $B$, $h_{i,j}$ is the channel gain from node $i$ to node $j$, and $x_k$ for $\forall k \in \{ p, s, b \}$ denotes the transmitted symbol from node $k$. The signal received at $D$ is then given by the following:
\begin{equation} 
    \begin{aligned}
    {y_{D}} &= \underbrace{\sqrt {P_s} {h_{s,d}}{x_{s}}}_{\text{$S$ desired}} + \underbrace{\sqrt {P_p} {h_{p,d}}{x_{p}}}_{\text{$P_{T}$ direct}} 
    + \underbrace{\sqrt {{\beta} P_p} {h_{p,b}}{h_{b,d}}{x_{p}}{x_{b}}}_{\text{$P_T$ backscatter}} \\
    &+ \underbrace{\sqrt {{\beta} P_s} {h_{s,b}}{h_{b,d}}{x_{s}}{x_{b}}}_{\text{$S$ backscatter}} + \underbrace{{n_{d}}}_{\text{noise}},
    \end{aligned}
\end{equation} 
where $n_d \sim \mathcal{CN}(0,\sigma_d^2)$ is the AWGN with zero mean and variance $\sigma_d^2$ at $D$. The received signal-to-interference-plus-noise ratio (SINR) at $P_R$ is given by:
\begin{equation} 
    {\gamma _{P}^p} = \frac {{{P_p}{{\left |{ {{h_{p,p}}} }\right |}^{2}}}} { { {{P_s} {\left |{ {{h_{s,p}}} }\right |}^{2}}  
    + \beta {\left |{ {{h_{b,p}}} }\right |}^{2} \chi_B + {\sigma _{p}^{2}}}}.
    \label{eq:gammap}
\end{equation}
where $\chi_B = \left( P_{s}{\left |{ {{h_{s,b}}} }\right |}^{2} + P_{p}{\left |{ {{h_{p,b}}} }\right |}^{2} \right)$.
The SINR for decoding $x_s$ at $D$ is given by
\begin{equation} 
    {\gamma _{D}^s} = \frac {{{P_s}{{\left |{ {{h_{s,d}}} }\right |}^{2}}}} { { {{P_p} {\left |{ {{h_{p,d}}} }\right |}^{2}}  
    + \beta {\left |{ {{h_{b,d}}} }\right |}^{2}  \chi_B  + {\sigma _{d}^{2}}}}.
\label{eq:gammas}    
\end{equation}
The SINR for decoding $x_b$ at $D$ with i-SIC is given by
\begin{equation} 
    {\gamma _{D}^b} = \frac {\beta{P_s}{{\left |{ {{h_{s,b}}} }\right |}^{2}{\left |{ {{h_{b,d}}} }\right |}^{2}}}
    { { {{P_p} {\left |{ {{h_{p,d}}} }\right |}^{2}} + \eta P_s |h_{s,d}|^2  
    + \beta P_{p}{\left |{ {{h_{p,b}}} }\right |}^{2}{\left |{ {{h_{b,d}}} }\right |}^{2}  + {\sigma _{d}^{2}}}},
    \label{eq:gammab}
\end{equation}
where $0 \leq \eta \leq 1$ is the i-SIC factor for decoding BD signal.
Other than $P_s$ and $\beta$, the SINR for decoding $x_b$ at $D$ is largely dominated by interference from $P_T$. 


\section{Joint Power-Reflection Optimization} \label{problem}
\subsection{Problem Formulation}

In this section, we formulate a joint optimization problem that maximizes the CR-AmBC data rates by jointly determining the optimal transmit power $P_s$ at $S$ and the reflection coefficient $\beta$ at $B$. The achievable data rates are given by
\begin{equation} 
    R_D^u(P_s,\beta) = \log_2 ( 1 + \gamma_D^u ) \ [\text{bits/s/Hz}], \ \ \forall u \in \{ s,b \}  
\end{equation}
where $R_D^u$ for all $\forall u \in \{ s, b \}$ represents the data rates for decoding $x_s$ and $x_b$. The main objective is to maximize the achievable data rates in the CRN without degrading the performance of the PN due to interference from both $S$ and $B$. The optimization problem is formulated as follows:
\stepcounter{equation}
\begin{align*} 
    & ({\mathbf{P}}1):\mathop {\max } \limits_{{\beta, P_s}} \ f(P_s,\beta) =\sum_{\forall u \in \{s,b\}}  \log_2 \left( 1 + \gamma_D^u \right)  \tag{\theequation} \label{eq:P1} \\ 
    & s.t.{\text{ }}{R_D^u}(P_s,\beta) \geq R_{u,\text{th}}, \ \  \forall{u} \in \{s,b\}\tag{C1} \\ 
    & \quad \quad P_s|h_{s,p}|^2 + \beta |h_{b,p}|^2 \chi_B   \leq {I_{p,\text{th}}},\tag{C2} \\
    & \quad \quad 0 \leq P_s \leq P_{\text{max}}, \tag{C3}  \\ 
    & \quad \quad 0 \leq \beta \leq   1,   \tag{C4}  
\end{align*}
where (C1) represents the minimum rates $R_{u,\text{th}}$ for $\forall u \in \{ s,b \}$ that are required to successfully decode $x_s$ and $x_b$, (C2) limits the interference below the threshold $I_{p,\text{th}}$, (C3) specifies the valid range for the transmit power $P_s$, and (C4) defines the reflection coefficient $\beta$. The optimization problem is further decomposed into two subproblems and reduced to a single dimensional search over $0 \leq \beta \leq 1$ given $P_s$. Algorithm~\ref{algo:main} presents the proposed solution based on the alternating optimization technique; for conciseness, the power and reflection-coefficient updates are given as Steps~1 and~2 of a single procedure, with the closed-form bounds and feasibility checks detailed in Sections~\ref{power} and~\ref{energy}.
%
\begin{algorithm}[tb!]
\caption{Alternating Optimization of Transmit Power and Reflection Coefficient for (P1)}
\label{alg:alternating-optimization}
\footnotesize
\begin{algorithmic}[1]
\Require \(|h_{i,j}|^2\), \(P_p\), \(\sigma_d^2\), \(I_{p,\text{th}}\), \(P_{\max}\), \(\eta\), \(\Gamma_{u,\text{th}} = 2^{R_{u,\text{th}}} - 1\) for \(u \in \{s,b\}\), grid \(\mathcal{B}=\{0,\frac{1}{N_\beta-1},\ldots,1\}\), \(\epsilon\), \(N_{\text{max}}\).
\State \textbf{Initialize:} \(\beta^{(0)} \in \mathcal{B}\), \(f^{(0)} \gets 0\), \texttt{flag}\(\gets 0\).
\For{iteration \(k = 1, 2, \ldots, N_{\text{max}}\)}
\Statex \quad \textit{Step 1 -- power update (P2), Section~\ref{power}:}
\State Compute \(P_{\min}(\beta^{(k-1)})\) by (\ref{eq:pmin})--(\ref{eq:pminb}) and \(P_{\max}(\beta^{(k-1)})\) by (\ref{eq:power}).
\If{conditions (\ref{eq:feasP}) hold}
  \State \(P_s^{(k)} \gets P_{\max}(\beta^{(k-1)})\) \Comment{optimal by Prop.~\ref{prop:power}}
\Else
  \State \Return \texttt{flag}\(\gets 1\) (infeasible/outage).
\EndIf
\Statex \quad \textit{Step 2 -- reflection update (P3), Section~\ref{energy}:}
\State Compute \(\beta_{\min}(P_s^{(k)})\) and \(\beta_{\text{upper}}(P_s^{(k)})\) by (\ref{eq:bs})--(\ref{eq:bmax}).
\If{\(\beta_{\min}(P_s^{(k)}) \leq \beta_{\text{upper}}(P_s^{(k)})\) and feasible}
  \State \(\beta^{(k)} \gets \arg\max_{\beta \in \mathcal{B}_k} f(P_s^{(k)},\beta)\), with \(\mathcal{B}_k = \big(\mathcal{B}\cap[\beta_{\min},\beta_{\text{upper}}]\big) \cup \{\beta_{\min},\beta_{\text{upper}},\beta^{(k-1)}\}\).
\Else
  \State \Return \texttt{flag}\(\gets 1\).
\EndIf
\State \(f^{(k)} \gets f(P_s^{(k)},\beta^{(k)})\); \textbf{if} \(|f^{(k)}-f^{(k-1)}| \leq \epsilon\) \textbf{then break}.
\EndFor
\State \textbf{Return:} \{\(P_s^* \gets P_s^{(k)}\), \(\beta^* \gets \beta^{(k)}\), \( f(P_s^*,\beta^*) \gets f^{(k)} \)\}.
\end{algorithmic}
\label{algo:main}
\end{algorithm}

\subsection{Power Allocation Optimization} \label{power}


The rates for both $D$ and $B$ increase with increasing $P_s$. For a fixed $\beta \in [0,1]$, the optimization problem to determine the optimal transmit power $P_s$ is formulated as follows:
\stepcounter{equation}
\begin{align*} 
    & ({\mathbf{P}}2):\mathop {\max } \limits_{{P_s}} \ f(P_s) =\sum_{\forall u \in \{s,b\}}  \log_2 \left( 1 + \gamma_D^u \right)  \tag{\theequation} \label{eq:P2} \\ 
    & \quad \quad \quad s.t.{\text{ (C1), (C2) and (C3) }}.
\end{align*}
For fixed $\beta$, the minimum limit for $P_s$ is given by
\begin{equation} 
    P_{\min}(\beta) = \max \left \{ P_{\min,s}(\beta),P_{\min,b}(\beta) \right \},
\label{eq:pmin}
\end{equation}
where $P_{\min,s}(\beta)$ and $P_{\min,b}(\beta)$ are the minimum transmit power levels required to meet the target rates for $x_s$ and $x_b$, respectively. Solving (C1) with (\ref{eq:gammas}) and (\ref{eq:gammab}) for $P_s$ gives
\begin{align}
    P_{\min,s}(\beta) &= \frac{\Gamma_{s,\text{th}} (P_p |h_{p,d}|^2 + \beta P_p |h_{p,b}|^2 |h_{b,d}|^2 + \sigma_d^2)}{|h_{s,d}|^2 - \Gamma_{s,\text{th}} \beta |h_{b,d}|^2 |h_{s,b}|^2}, \label{eq:pmins}\\
    P_{\min,b}(\beta) &= \frac{\Gamma_{b,\text{th}} (P_p |h_{p,d}|^2 + \beta P_p |h_{p,b}|^2 |h_{b,d}|^2 + \sigma_d^2)}{\beta |h_{s,b}|^2 |h_{b,d}|^2 - \Gamma_{b,\text{th}} \eta |h_{s,d}|^2}, \label{eq:pminb}
\end{align}
which are valid when the respective denominators are positive. The solution for the maximum allowable transmit power as a function of a fixed $\beta$ is determined based on the interference threshold $I_{p,\text{th}}$, given by
\begin{equation}
    P_s(\beta) \leq \min \left \{ P_{\max},\frac{I_{p,\text{th}} - \beta|h_{b,p}|^2 P_p|h_{p,b}|^2  }
                   {|h_{s,p}|^2 + \beta|h_{s,b}|^2|h_{b,p}|^2} \right \} \triangleq P_{\max}(\beta).
\label{eq:power}
\end{equation}
The maximum allowable transmit power at $S$ is mainly limited by the interference threshold $I_{p,\text{th}}$ at the PN receiver. Therefore, based on (\ref{eq:power}), the following conditions must hold for a feasible solution: ${P_{\min}(\beta) \leq P_s}(\beta)$ and ${I_{p,\text{th}} \geq \beta |h_{b,p}|^2 P_p |h_{p,b}|^2}$; otherwise, $P_s(\beta) = 0$. Collecting these requirements, (P2) is feasible if and only if
\begin{equation}
\begin{aligned}
    &|h_{s,d}|^2 > \Gamma_{s,\text{th}}\beta |h_{b,d}|^2 |h_{s,b}|^2, \ \ \beta |h_{s,b}|^2 |h_{b,d}|^2 > \Gamma_{b,\text{th}} \eta |h_{s,d}|^2, \\
    &I_{p,\text{th}} \geq \beta |h_{b,p}|^2 P_p |h_{p,b}|^2, \ \ \text{and} \ \ P_{\min}(\beta) \leq P_{\max}(\beta).
\end{aligned}
\label{eq:feasP}
\end{equation}
\begin{proposition}\label{prop:power}
For a fixed $\beta \in (0,1]$ satisfying (\ref{eq:feasP}), the unique optimal solution of (P2) is $P_s^*(\beta) = P_{\max}(\beta)$.
\end{proposition}
\begin{IEEEproof}
For fixed $\beta$, (\ref{eq:gammas}) and (\ref{eq:gammab}) have the form $\gamma_D^u = a_u P_s/(b_u P_s + c_u)$ with $a_u>0$, $b_u \geq 0$, $c_u \geq \sigma_d^2 > 0$, so $\partial \gamma_D^u/\partial P_s = a_u c_u/(b_u P_s + c_u)^2 > 0$ and $f$ is strictly increasing in $P_s$. Since $\gamma_D^u < a_u/b_u$, (C1) requires $a_u > \Gamma_{u,\text{th}} b_u$ (the first two conditions in (\ref{eq:feasP})) and is then equivalent to $P_s \geq P_{\min}(\beta)$, while (C2)--(C3) are equivalent to $P_s \leq P_{\max}(\beta)$. The maximum of $f$ over $[P_{\min}(\beta), P_{\max}(\beta)]$ is thus uniquely attained at the right end point.
\end{IEEEproof}
Step~1 of Algorithm~\ref{algo:main} implements this closed-form update.

\subsection{Reflection Coefficient Optimization} \label{energy}

For a fixed $P_s$, the optimal reflection coefficient $\beta \in [0,1]$ that maximizes $f(\beta)$ can be obtained via numerical evaluation over the interval [0,1]. For example, the upper bound based on the BD constraint can be evaluated for a feasibility check at $\beta = 1$, such that $\gamma_D^b(\beta \gets 1) \geq \Gamma_{b,\text{th}} = (2^{R_{b,\text{th}}}-1)$. The optimization problem is formulated as follows:
\stepcounter{equation}
\begin{align*} 
    & ({\mathbf{P}}3):\mathop {\max } \limits_{{\beta}} \ f(\beta) =\sum_{\forall u \in \{s,b\}}  \log_2 \left( 1 + \gamma_D^u \right)  \tag{\theequation} \label{eq:P3} \\ 
    & \quad \quad \quad s.t. {\text{ (C1), (C2) and (C4). }} 
\end{align*}
From (C1), the solution for $\beta$ must satisfy the following:
\begin{equation} 
    \beta \leq \beta_{\min,s}(P_s) \gets \frac{P_s |h_{s,d}|^2 - \Gamma_{s,\text{th}} (P_p |h_{p,d}|^2 + \sigma_d^2)}{\Gamma_{s,\text{th}} |h_{b,d}|^2 (P_s |h_{s,b}|^2 + P_p |h_{p,b}|^2)}, 
\label{eq:bs}
\end{equation}
where $\Gamma_{s,\text{th}} = (2^{R_{s,{\text{th}}}}-1)$ is derived from the rate constraint $R_{s,{\text{th}}}$, based on which the feasibility check includes $P_s |h_{s,d}|^2 > \Gamma_{s,\text{th}} (P_p |h_{p,d}|^2 + \sigma_d^2)$. Also, the following must be satisfied for decoding $x_b$:
\begin{equation} 
    \beta \geq \beta_{\min,b}(P_s) \gets \frac{\Gamma_{b,\text{th}} (P_p |h_{p,d}|^2 + \eta P_s |h_{s,d}|^2 + \sigma_d^2)}{P_s |h_{s,b}|^2 |h_{b,d}|^2 - \Gamma_{b,\text{th}} P_p |h_{p,b}|^2 |h_{b,d}|^2},
\label{eq:bb}
\end{equation}
and the feasibility check must satisfy the condition that $P_s |h_{s,b}|^2 |h_{b,d}|^2 > \Gamma_{b,\text{th}} P_p |h_{p,b}|^2 |h_{b,d}|^2$. From (C2), the following must hold to limit the interference to $P_R$:
\begin{equation} 
    \beta \leq \beta_{\max}(P_s) \gets \frac{I_{p,\text{th}} - P_s |h_{s,p}|^2}{|h_{b,p}|^2 (P_s |h_{s,b}|^2 + P_p |h_{p,b}|^2)}.
\label{eq:bmax}
\end{equation}
The upper bound for the feasible range of $\beta$ is therefore defined by $\beta_{\text{upper}}(P_s) \gets \min \{\beta_{\min,s}(P_s),\beta_{\max}(P_s),1\}$, and the lower bound by $\beta_{\min} \gets \max \left\{0, \beta_{\min,b}(P_s) \right\}$, where (\ref{eq:bmax}) additionally requires $P_s |h_{s,p}|^2 \leq I_{p,\text{th}}$. Step~2 of Algorithm~\ref{algo:main} provides further details on the solution.

\begin{remark}\label{rem:shape}
For fixed $P_s$, $R_D^s = \log_2(1+a/(b\beta+c))$ is convex and decreasing in $\beta$, whereas $R_D^b = \log_2(1+d\beta/(e\beta+g))$ is concave and increasing, with $a,\ldots,g>0$ given by (\ref{eq:gammas})--(\ref{eq:gammab}). Hence $\beta$ trades one link against the other (and the PN through (C2)), and $f(P_s,\cdot)$ is not guaranteed to be unimodal, which is why Step~2 uses a grid search rather than a golden-section search.
\end{remark}

\begin{remark}\label{rem:eh}
A BD energy-harvesting constraint $\zeta(1-\beta)\chi_B \geq P_c$, with efficiency $\zeta$ and circuit power $P_c$, adds the bound $\beta \leq 1 - P_c/(\zeta\chi_B)$ to $\beta_{\text{upper}}(P_s)$ and a linear lower bound on $P_s$ to $P_{\min}(\beta)$, so Algorithm~\ref{algo:main} and Proposition~\ref{prop:power} still apply.
\end{remark}


\subsection{Monotonicity and Convergence of Alternating Updates}
The proposed solution updates \(P_s\) using Step~1 of Algorithm~\ref{algo:main}, and \(\beta\) using Step~2 of Algorithm~\ref{algo:main} in a block-coordinate procedure. The following summarizes the convergence properties:

\begin{proposition}\label{prop:conv}
Let \(f(P_s,\beta)=\sum_{u\in\{s,b\}} \log_2(1+\gamma_D^u)\) be the objective and assume the feasible set defined by the constraints (C1)--(C4) is nonempty and compact. Then each alternating update does not decrease the objective value (i.e., the sequence \(\{f(P_s^{(k)},\beta^{(k)})\}_k\) is non-decreasing). Moreover, the sequence converges, and for \(\epsilon>0\) Algorithm~\ref{algo:main} terminates after at most \(\lceil (\bar{f}-f^{(1)})/\epsilon \rceil + 1\) iterations, where \(\bar{f} = \log_2(1+P_{\max}|h_{s,d}|^2/\sigma_d^2) + \log_2(1+P_{\max}|h_{s,b}|^2|h_{b,d}|^2/\sigma_d^2)\).
\end{proposition}

\begin{IEEEproof}
Step~1 is optimal for fixed \(\beta\) (Proposition~\ref{prop:power}) and Step~2 is optimal over \(\mathcal{B}_k \ni \beta^{(k-1)}\) for fixed \(P_s\), so \(f\) is non-decreasing. Since \(\gamma_D^s \leq P_{\max}|h_{s,d}|^2/\sigma_d^2\) and \(\gamma_D^b \leq P_{\max}|h_{s,b}|^2|h_{b,d}|^2/\sigma_d^2\), \(f \leq \bar{f}\), and each non-terminating iteration increases \(f\) by more than \(\epsilon\).
\end{IEEEproof}

Proposition~\ref{prop:power} further allows the joint problem to be characterized exactly.

\begin{corollary}\label{cor:1d}
Let \(\mathcal{B}_f \subseteq (0,1]\) be the set of \(\beta\) satisfying (\ref{eq:feasP}) and \(g(\beta) \triangleq f(P_{\max}(\beta),\beta)\). Then (P1) is equivalent to the one-dimensional problem \(\max_{\beta \in \mathcal{B}_f} g(\beta)\).
\end{corollary}
\begin{IEEEproof}
By Proposition~\ref{prop:power}, \(f(P_s,\beta) \leq g(\beta)\) for any feasible \((P_s,\beta)\), with equality at the feasible point \((P_{\max}(\beta),\beta)\).
\end{IEEEproof}

Hence, initializing \(\beta^{(0)}\) with the grid maximizer of the continuous \(g\) makes Algorithm~\ref{algo:main} at least as good as the best grid point of (P1).

\subsection{Computational Complexity}\label{complexity}
For each alternating iteration, the update for $P_s$ requires closed-form feasibility checks and evaluation of algebraic expressions in constant time, $\mathcal{O}(1)$, namely (\ref{eq:pmin})--(\ref{eq:feasP}). For \(\beta\), the bounds (\ref{eq:bs})--(\ref{eq:bmax}) cost $\mathcal{O}(1)$ and \(f\) is evaluated at \(|\mathcal{B}_k| \leq N_\beta+3\) points, i.e., $\mathcal{O}(N_\beta)$. The overall complexity of Algorithm~\ref{algo:main} is therefore $\mathcal{O}(KN_\beta)$, with $K$ bounded by Proposition~\ref{prop:conv}, versus $\mathcal{O}(N_PN_\beta)$ for an exhaustive search over $N_P$ power levels and an interior-point convex program per iteration for SDR/SCA designs \cite{Xu2024ris}, \cite{Rezaei2024noma}. With $N_\beta = 100$, an iteration requires about $10^3$ floating-point operations, negligible relative to the channel coherence time.

\section{Simulation Results and Discussion} \label{results}

\begin{table}[bp!]
\caption{Simulation Parameters}
\begin{center}
\begin{tabular}{l|l}
\hline
\textbf{Parameter} & \textbf{Value} \\
\hline \hline
Monte Carlo simulations & $10^{5} ~\text{iterations}$  \\
PU transmit power & $P_p = 30 ~\text{dBm}$  \\
CR-AmBC transmit power & $P_s = [0,P\max]$  \\
Reflection coefficient & $\beta = [0,1]$  \\
Interference threshold at $P_R$ & $ I_{p,\text{th}} = 10 ~\text{dBm} $ \\
PU minimum data rate & $ R_{p,\text{th}} = 1 \ \text{[bit/s/Hz]} $ \\
CRN minimum data rate & $ R_{s,\text{th}} = 0.5 \ \text{[bits/s/Hz]} $ \\
BD minimum data rate & $ R_{b,\text{th}} = 0.1 \ \text{[bits/s/Hz]} $ \\
Noise power & $\sigma^2 = -30 ~\text{dBm}$  \\
Path loss & $\lambda_{i,j} = d_{i,j}^{-\alpha}$  \\
Path loss exponent & $\alpha = 2$  \\
i-SIC & $\eta = [0.1,0.4,0.6]$  \\
\hline
\end{tabular}
\label{tab:params}
\end{center}
\end{table}


\begin{figure}[!b]
  \centering
  \includegraphics[width=0.85\columnwidth]{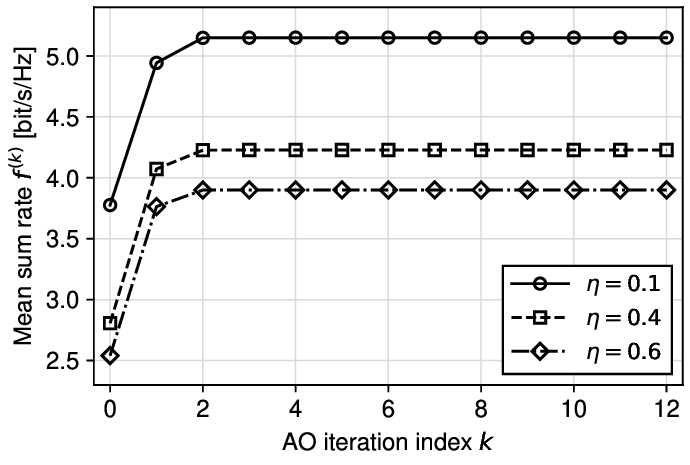}
  \caption{Convergence of Algorithm~\ref{algo:main} for different i-SIC factors ($P_{\max} = 30$~dBm, $P_p = 30$~dBm, $I_{p,\text{th}} = 10$~dBm).}
  \label{fig:conv}
\end{figure}

\begin{figure*}[t]
  \centering
  \begin{minipage}{0.30\textwidth}
    \centering
    \includegraphics[width=\linewidth]{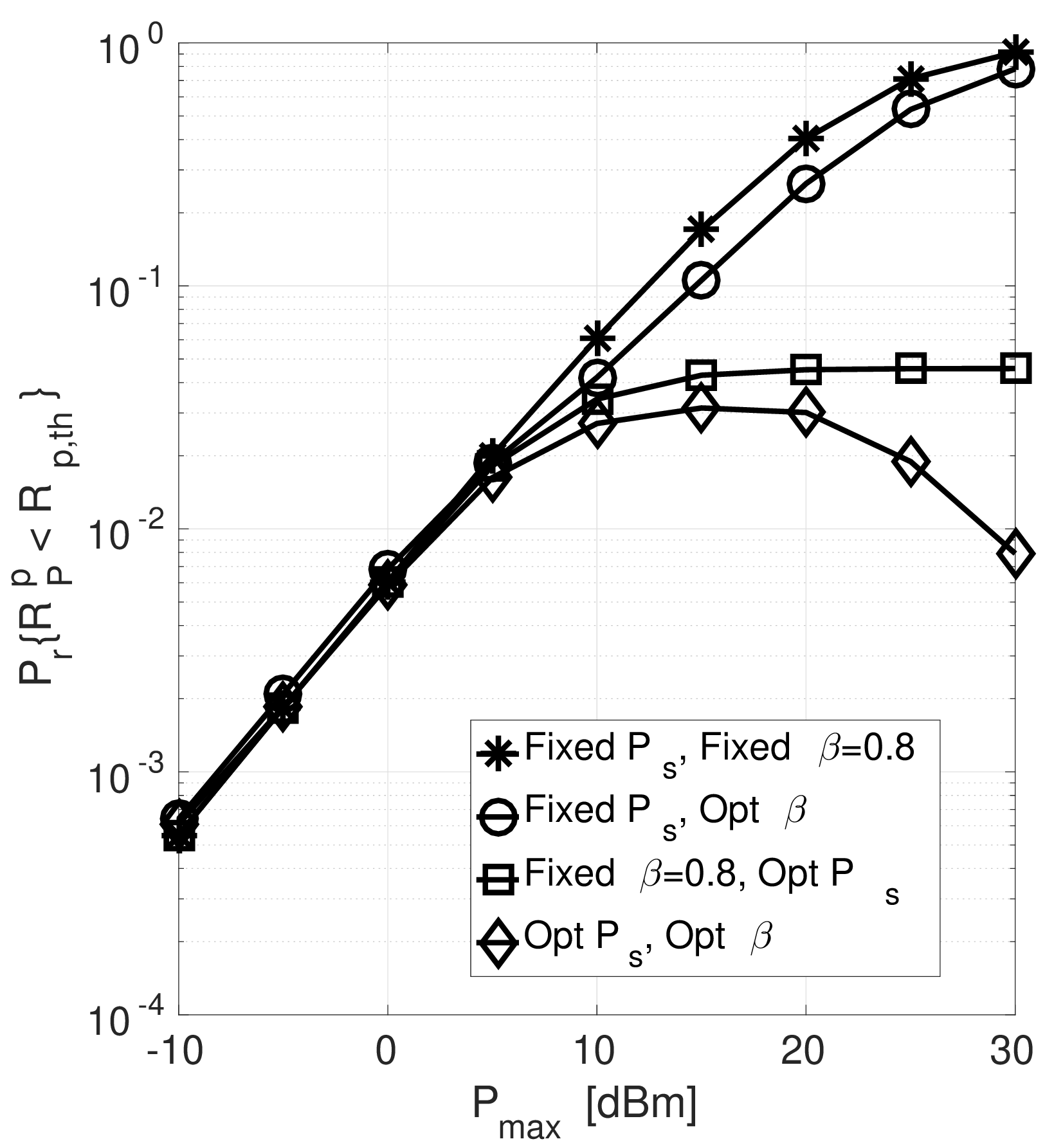}
    \caption{PU outage over CRN transmit power ($P_p = 30$~dBm, $I_{p,\text{th}} = 10$~dBm).}
    \label{fig:fig002}
  \end{minipage}\hfill
  \begin{minipage}{0.30\textwidth}
    \centering
    \includegraphics[width=\linewidth]{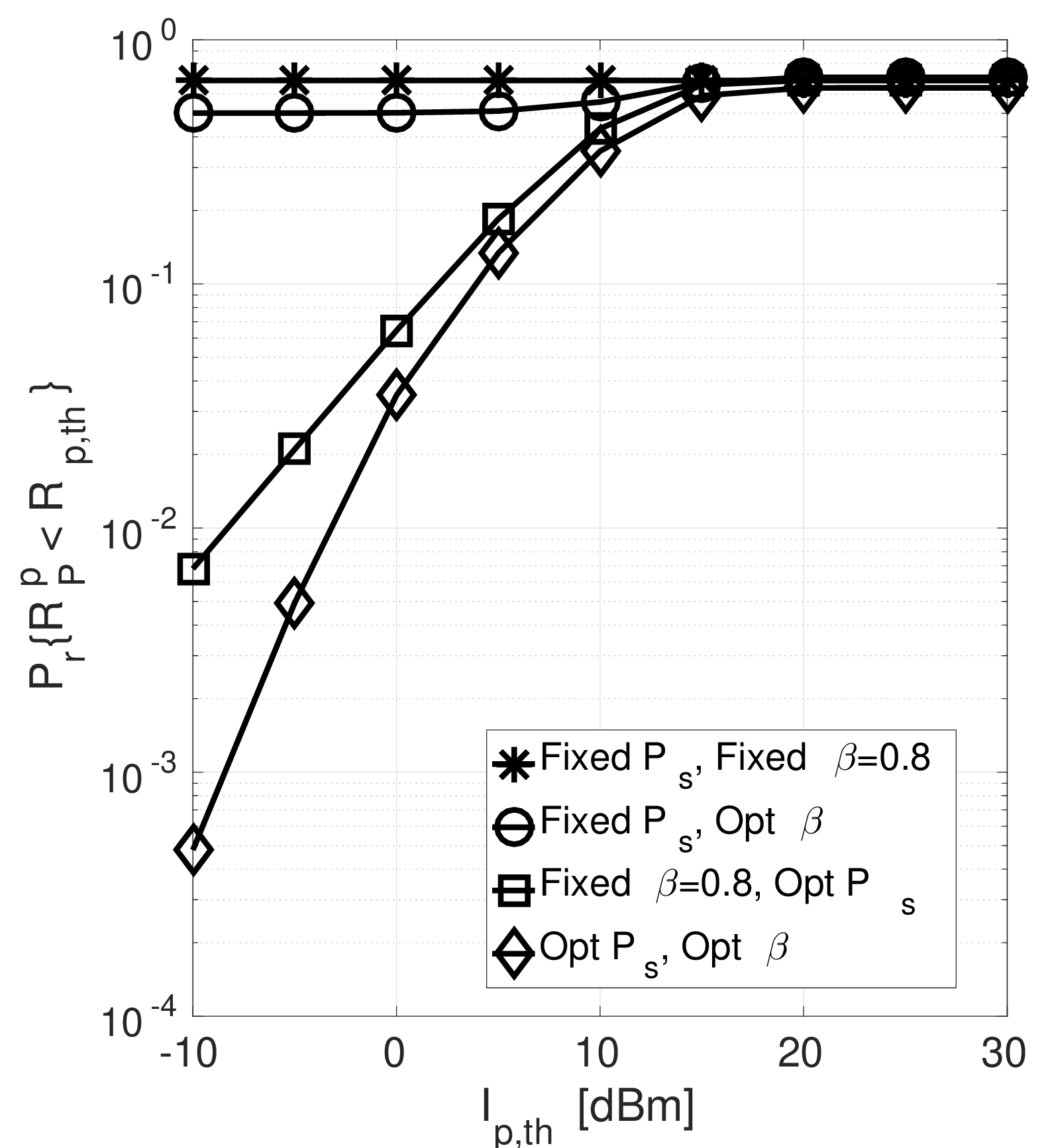}
    \caption{PU outage over interference threshold $I_{p,\text{th}}$ ($P_p = 30$~dBm).}
    \label{fig:fig003}
  \end{minipage}\hfill
  \begin{minipage}{0.30\textwidth}
    \centering
    \includegraphics[width=\linewidth]{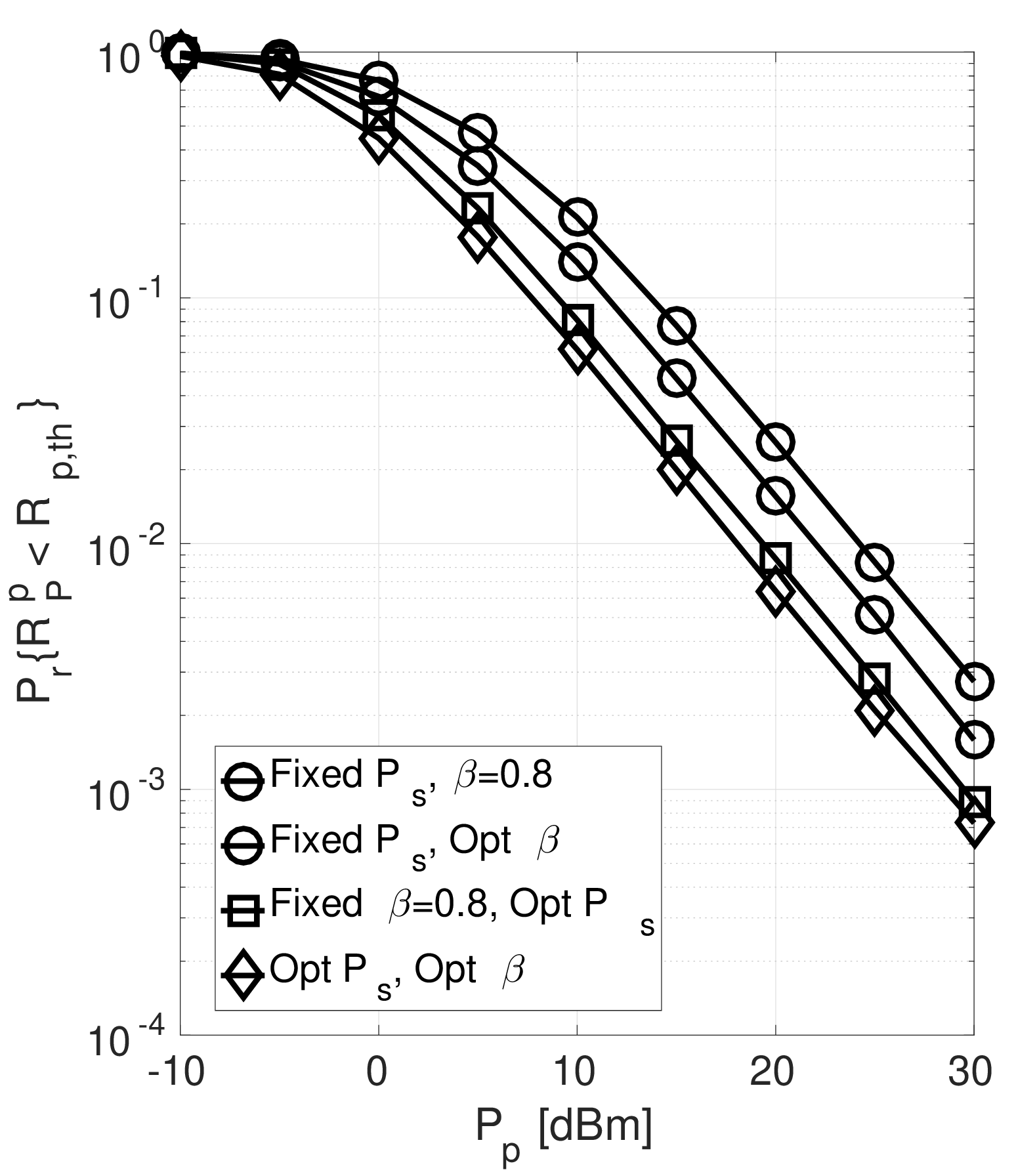}
    \caption{PU outage over PN transmit power $P_p$ ($I_{p,\text{th}} = 10$~dBm).}
    \label{fig:fig004}
  \end{minipage}\hfill
\end{figure*}


\begin{figure*}[t]
  \centering
  \begin{minipage}{0.30\textwidth}
    \centering
    \includegraphics[width=\linewidth]{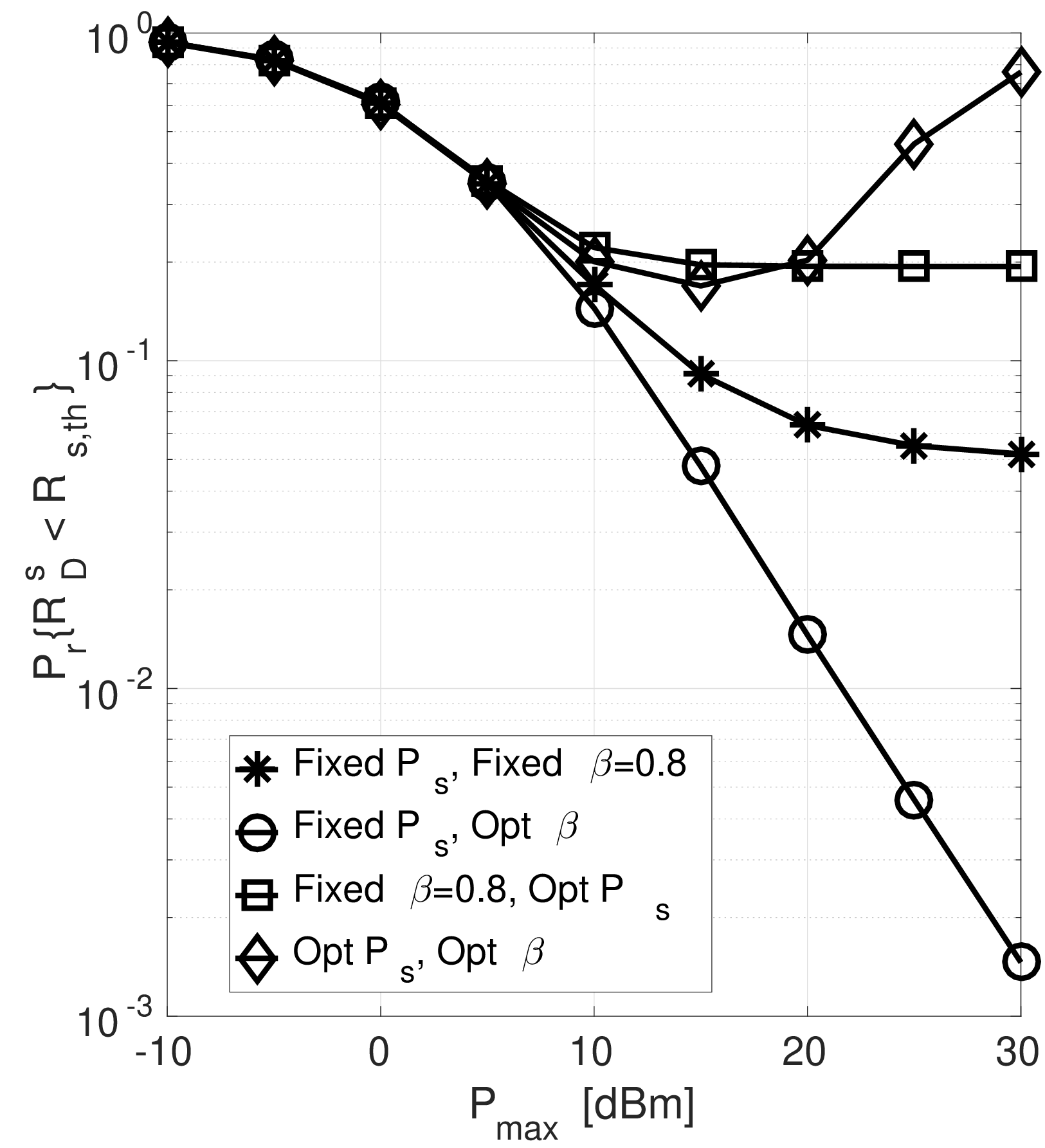}
    \caption{CRN outage in decoding $x_s$ over CRN transmit power ($P_p = 30$~dBm, $I_{p,\text{th}} = 10$~dBm).}
    \label{fig:fig005}
  \end{minipage}\hfill
  \begin{minipage}{0.30\textwidth}
    \centering
    \includegraphics[width=\linewidth]{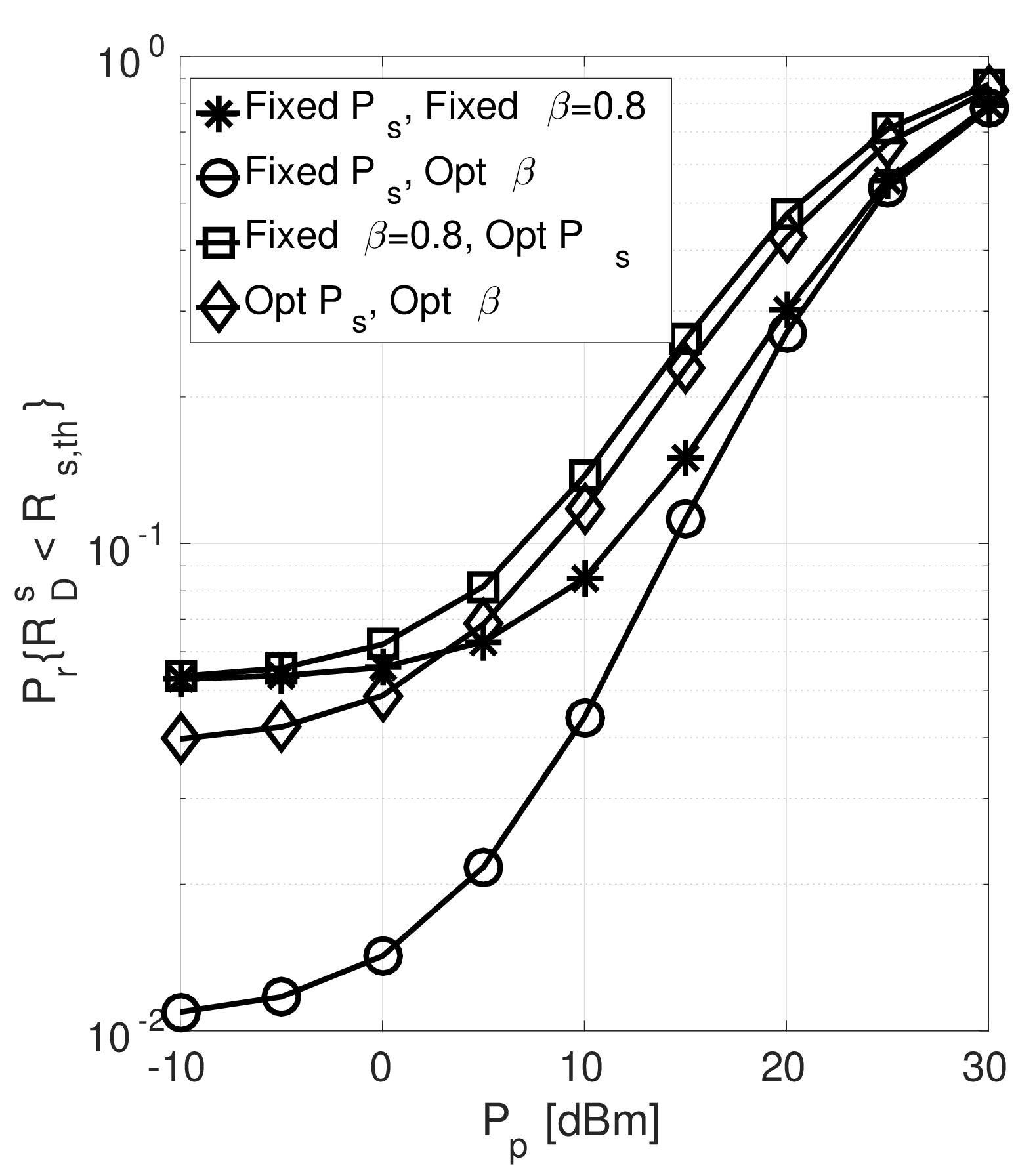}
    \caption{CRN outage in decoding $x_s$ over PN transmit power $P_p$ ($I_{p,\text{th}} = 10$~dBm).}
    \label{fig:fig006}
  \end{minipage}\hfill
  \begin{minipage}{0.30\textwidth}
    \centering
    \includegraphics[width=\linewidth]{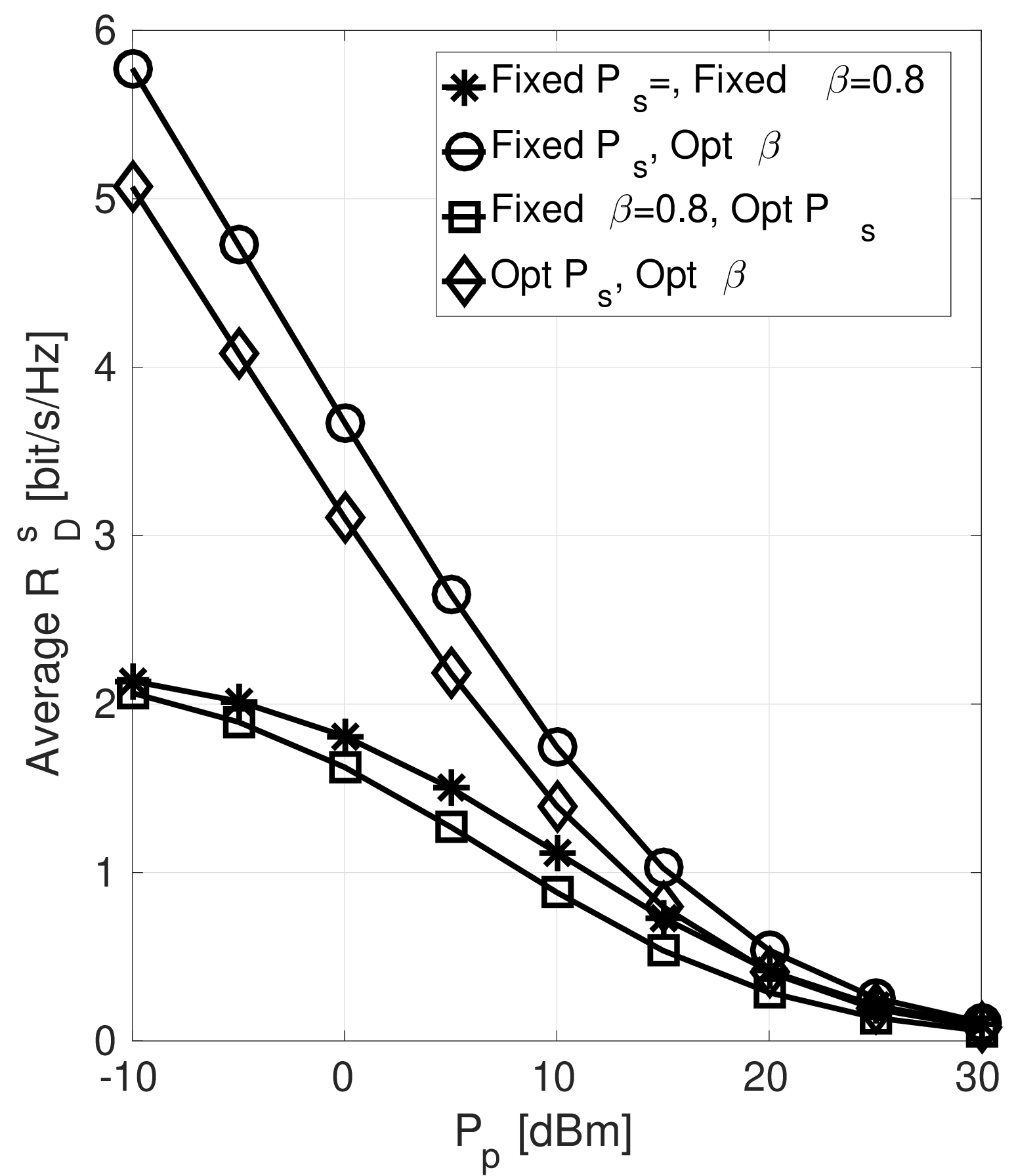}
    \caption{Average CRN data rate $R_D^s$ over PN transmit power $P_p$ ($I_{p,\text{th}} = 10$~dBm).}
    \label{fig:fig007}
  \end{minipage}\hfill
\end{figure*}



This section presents Monte Carlo simulation results over $10^5$ iterations. The simulation parameter settings are provided in Table \ref{tab:params}. The results investigate various factors affecting performance at $P_R$ with respect to $x_p$, and at $D$ with respect to $x_s$ and $x_b$. These factors include the available transmit power $P_{\text{max}}$ at $S$, the transmit power of $P_T$, the interference threshold $I_{p,\text{th}}$ at $P_R$, and the impact of i-SIC on decoding $x_b$ at $D$. The proposed scheme (Opt $P_s$, Opt $\beta$) is compared with: (a) fixed $P_s = P_{\max}$, $\beta = 0.8$; (b) fixed $P_s = P_{\max}$ with $\beta$ from Step~2, emulating the reflection-coefficient-only designs of \cite{Thenua2024}, \cite{Zhuang2020}; and (c) fixed $\beta = 0.8$ with $P_s$ from Step~1, emulating the adaptive-power designs of \cite{Badarneh2022}, \cite{Zhang2023}.


\begin{figure*}[t]
  \centering
  \begin{minipage}{0.30\textwidth}
    \centering
    \includegraphics[width=\linewidth]{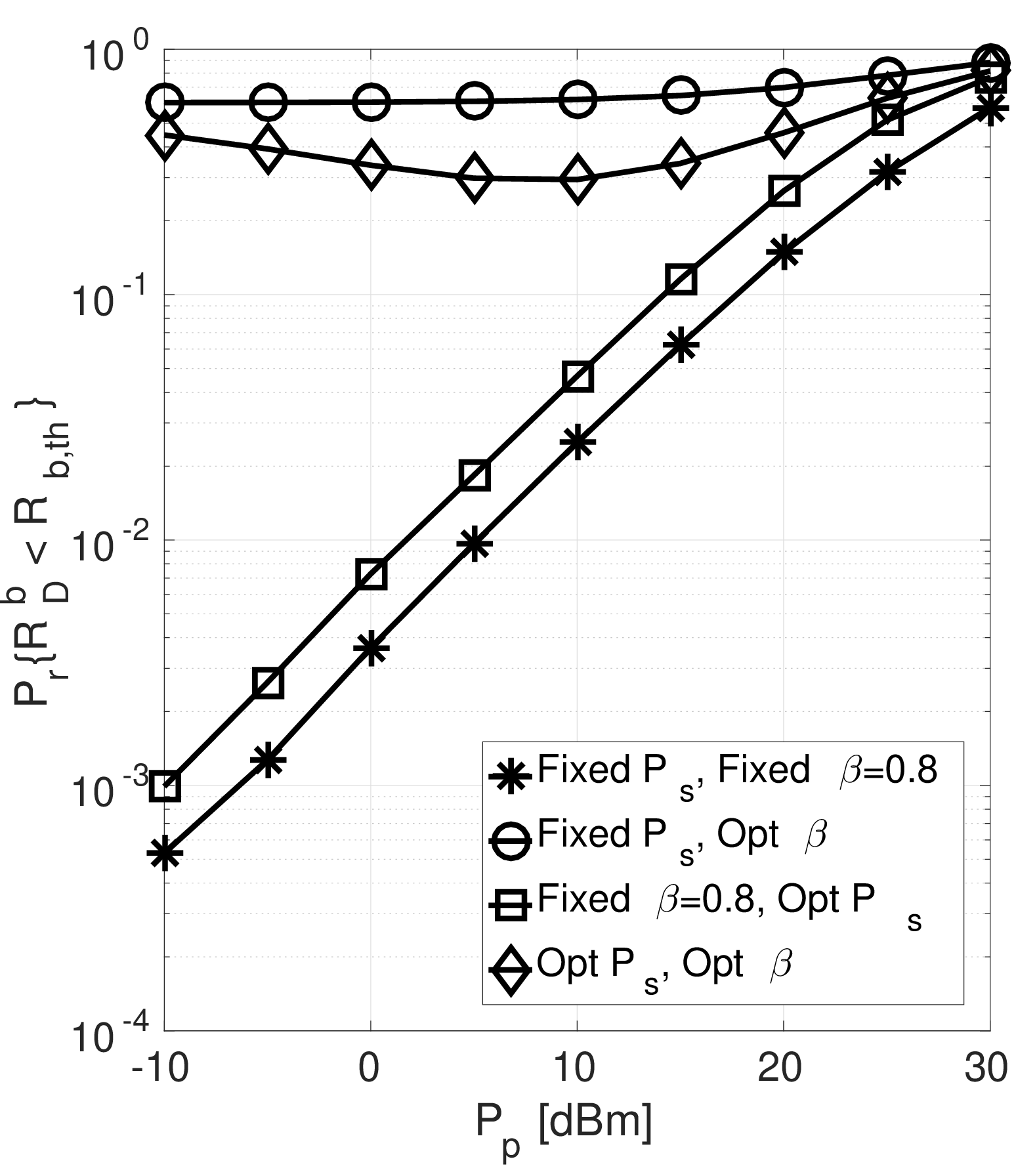}
    \caption{BD outage in decoding $x_b$ over PN transmit power $P_p$ ($I_{p,\text{th}} = 10$~dBm).}
    \label{fig:fig008}
  \end{minipage}\hfill
  \begin{minipage}{0.30\textwidth}
    \centering
    \includegraphics[width=\linewidth]{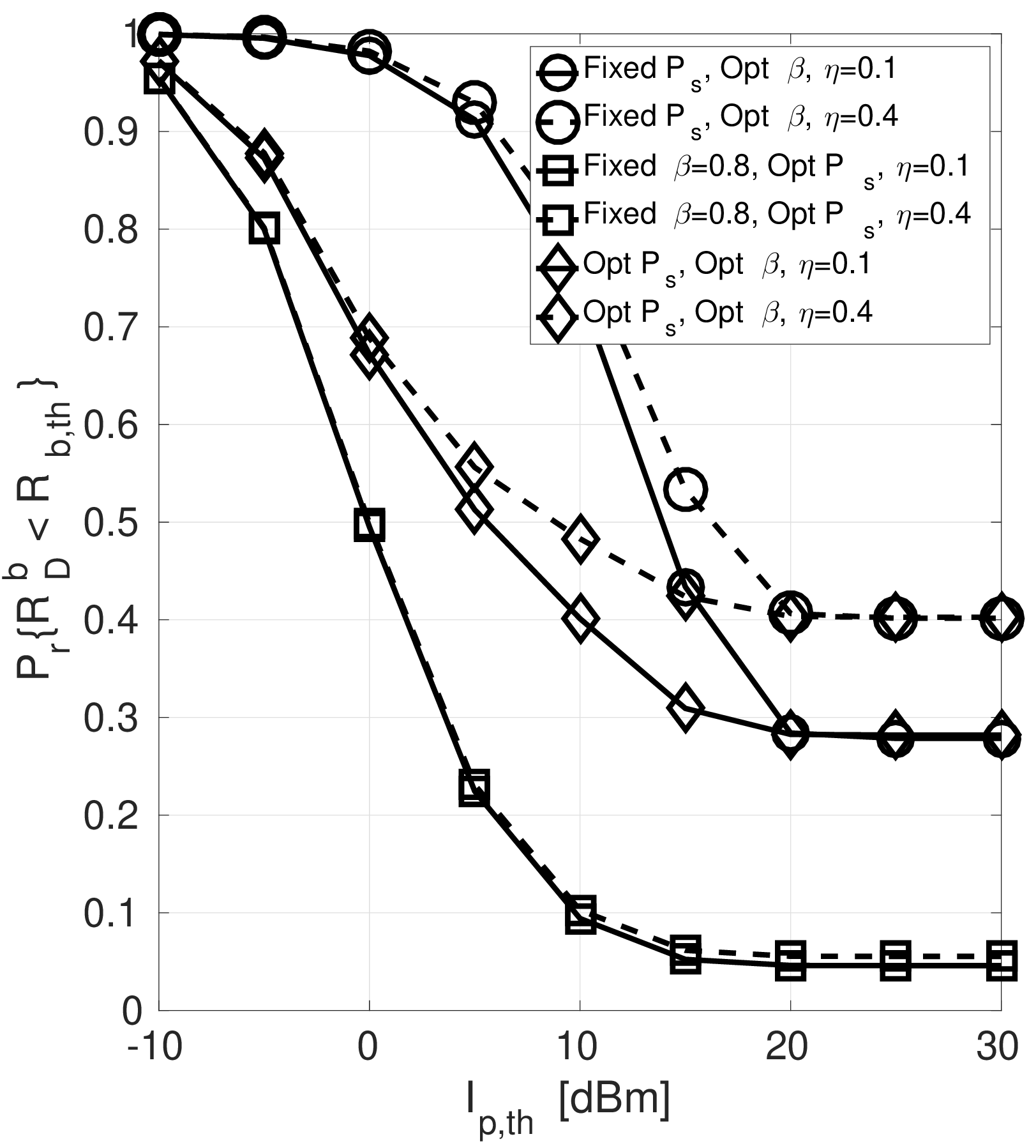}
    \caption{BD outage over PN interference threshold $I_{p,\text{th}}$ with i-SIC factors $\eta \in \{0.1, 0.4\}$ ($P_p = 30$~dBm).}
    \label{fig:fig009}
  \end{minipage}\hfill
  \begin{minipage}{0.30\textwidth}
    \centering
    \includegraphics[width=\linewidth]{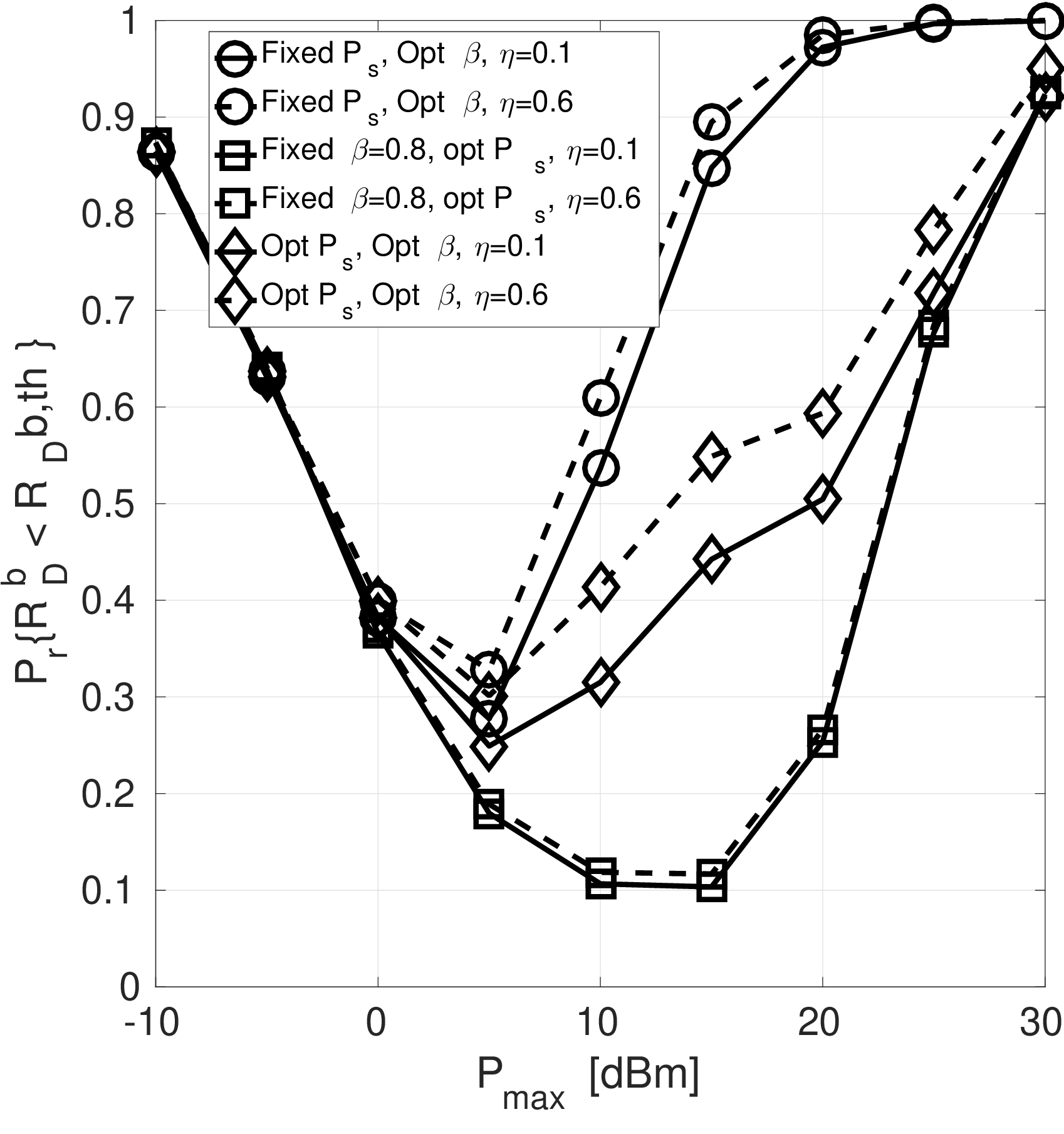}
    \caption{BD outage over CRN transmit power with i-SIC factors $\eta \in \{0.1, 0.6\}$ ($P_p = 30$~dBm, $I_{p,\text{th}} = 10$~dBm).}
    \label{fig:fig010}
  \end{minipage}\hfill

\end{figure*}


Fig.~\ref{fig:conv} verifies the convergence of Algorithm~\ref{algo:main} ($N_\beta = 100$, $\epsilon = 10^{-4}$, $\beta^{(0)} = 0.5$, $2\times10^4$ channel draws): the objective is non-decreasing in every realization (Proposition~\ref{prop:conv}), about 97\% of the realizations stop within four iterations for all $\eta$, and the solution attains 98.8--99.8\% of the exhaustive-search optimum (above 99.95\% with the warm start of Corollary~\ref{cor:1d}).

In Figs. \ref{fig:fig002} to \ref{fig:fig004}, we present the PN outage performance, which evaluates the probability that the data rate $R_P^p$ at $P_R$ falls below a predefined threshold $R_{p,\text{th}}$, i.e., $\Pr\{R_P^p < R_{p,\text{th}}\}$. Fig. \ref{fig:fig002} shows that increasing the transmit power $P_s$ increases the PU outage; however, the joint optimization of transmit power $P_s$ and reflection coefficient $\beta$ is effective in reducing the outage. Fig. \ref{fig:fig003} illustrates that limiting the interference at $P_R$ reduces PU outage. Moreover, Fig. \ref{fig:fig004} shows that increasing the PU transmit power $P_p$ decreases outage as expected, further highlighting the effectiveness of jointly optimizing $P_s$ and $\beta$ in reducing PU outage. In Figs. \ref{fig:fig005} to \ref{fig:fig007}, we present the results for the performance of the CRN in decoding $x_s$ at $D$. Fig. \ref{fig:fig005} shows that, for the proposed scheme with $P_{\max}$ above approximately 15~dBm, increasing the transmit power $P_s$ at $S$ leads to higher outage in decoding $x_s$. This increase is due to the reflected power from the BD. Indeed, from (\ref{eq:gammas}) and (\ref{eq:gammab}), as $P_s$ grows, $\gamma_D^s \to |h_{s,d}|^2/(\beta|h_{b,d}|^2|h_{s,b}|^2)$ and $\gamma_D^b \to \beta|h_{s,b}|^2|h_{b,d}|^2/(\eta|h_{s,d}|^2)$, so that the product of the two SINR ceilings equals $1/\eta$. Increasing $P_s$ thus cannot remove the outage floors caused by the self-backscatter loop $S \to B \to D$ and the residual SIC error, which explains the saturation of the fixed-$P_s$, fixed-$\beta$ curve in Fig.~\ref{fig:fig005}. Figs. \ref{fig:fig006} and \ref{fig:fig007} illustrate that increasing $P_p$ further degrades the performance of decoding $x_s$. Nonetheless, performance can be significantly improved by optimizing $\beta$ to reduce interference caused by reflected signals from both $P_T$ and $S$. Therefore, if the reflected signals from $P_T$ and $S$ are not effectively controlled through $\beta$, the decoding performance of $x_s$ at $D$ is reduced. 

In Figs. \ref{fig:fig008} to \ref{fig:fig010}, we present the results for decoding the BD signal $x_b$ at $D$. The decoding of $x_b$ is also subject to imperfect SIC (i-SIC), characterized by the factor $\eta$. Fig. \ref{fig:fig008} illustrates the impact of $P_p$, showing that at high levels of $P_p$, neither optimizing $P_s$ nor $\beta$ can sufficiently mitigate the interference from $P_T$. Fig. \ref{fig:fig009} presents the impact of varying $I_{p,\text{th}}$ on $x_b$, indicating that relaxing the PU interference constraint improves the performance of the BD signal. Fig. \ref{fig:fig010} demonstrates that there exists an optimal level of $P_s$ that minimizes the outage probability for the BD; increasing $P_s$ improves performance up to this optimal level, beyond which performance deteriorates. This occurs because both $P_s$ and $\beta$ must be jointly optimized to protect $P_R$ from interference. Furthermore, Figs. \ref{fig:fig009} and \ref{fig:fig010} show that a high i-SIC factor reduces BD performance, highlighting the critical importance of accurate SIC in decoding the BD signal. The baselines with fixed $\beta = 0.8$ achieve a lower BD outage at the cost of a higher PN outage and a lower CRN rate (Figs.~\ref{fig:fig002}--\ref{fig:fig004}, \ref{fig:fig007}), reflecting the trade-off of Remark~\ref{rem:shape}; a weighted sum rate can prioritize the BD without changing Algorithm~\ref{algo:main}.

For the settings of Table~\ref{tab:params}, (i) a power budget of about 5--15~dBm minimizes the BD outage (Fig.~\ref{fig:fig010}); (ii) relaxing $I_{p,\text{th}}$ beyond about 15--20~dBm brings no BD gain but raises the PN outage (Figs.~\ref{fig:fig003}, \ref{fig:fig009}); and (iii) since $\gamma_D^s\gamma_D^b \to 1/\eta$, SIC quality matters more than transmit power.

\balance

\section{Conclusions} \label{conclutions}

In this paper, we present an underlay CR-AmBC system operating under a PN interference constraint. We propose an optimization framework to determine the optimal transmit power and reflection coefficient. Simulation results demonstrate that interference across both the PN and CR-AmBC networks significantly impacts system performance. Furthermore, the effect of reflected interference signals in both networks can be minimized through joint optimization of transmit power and reflection coefficient. The proposed joint optimization framework enhances PN reliability while simultaneously improving the performance of the CR-AmBC system. These results highlight the importance of considering interference from both the PN and CR-AmBC in system design and provide valuable insights for the development of energy-efficient and spectrum-efficient IoT communications. The optimal closed-form power update reduces the problem to a one-dimensional search, and the algorithm converges within a few iterations with $\mathcal{O}(KN_\beta)$ complexity. Future work will address multiple BDs \cite{Thenua2024}, multi-antenna nodes \cite{Rezaei2024noma}, imperfect CSI \cite{Xu2024ris}, and comparisons with existing designs under a common model.

\bibliographystyle{IEEEtran}
\bibliography{references}

\end{document}